\documentclass[usenatbib]{mnras}
\usepackage[utf8]{inputenc}
\usepackage{natbib,epsfig,cite,amsmath,amssymb,graphicx}
\usepackage{array, xcolor, caption, multirow}

\def\e10{\eta_{10}}

\def\iso#1#2{\mbox{${}^{#2}{\rm #1}$}}

\newcommand\li[1]{\iso{Li}{#1}}
\def\be#1{\iso{Be}{#1}}
\def\li#1{\iso{Li}{#1}}
\def\b1#1{\iso{B}{1#1}}

\def\beq{\begin{equation}}
\def\eeq{\end{equation}}
\def\beqar{\begin{eqnarray}}
\def\eeqar{\end{eqnarray}}
\def\simlt{\lower.5ex\hbox{$\; \buildrel < \over \sim \;$}}
\def\simgt{\lower.5ex\hbox{$\; \buildrel > \over \sim \;$}}
\def\simpropto{\lower.2ex\hbox{$\; \buildrel \propto \over \sim \;$}}

\title[$\nu$-Process F and CR Nucleosynthesis]{Constraining $\nu$-Process Production of Fluorine through Cosmic Ray Nucleosynthesis}
 
\author[K.A. Olive and E. Vangioni]
{Keith A. Olive$^{1}$\thanks{e-mail:olive@umn.edu},
Elisabeth Vangioni$^{2}$\thanks{e-mail:vangioni@iap.fr}\\
$^{1}$William I. Fine Theoretical Physics Institute, School of Physics and Astronomy, 
University of Minnesota, Minneapolis, MN 55455, USA\\
$^{2}$Sorbonne Universit\'e, UPMC Univ Paris 6 et CNRS, UMR 7095, Institut d'Astrophysique de Paris, 98 bis bd Arago, 75014 Paris, France\\
}
 
\begin{document}

\pagerange{\pageref{firstpage}--\pageref{lastpage}} \pubyear{2018}
\maketitle
\label{firstpage}

\begin{abstract}
Fluorine is massive enough that it is not considered to be a light ($Z\le5$) element, yet compared to its near neighbors,
C, N, O,  and Ne, it is far underproduced in the course of stellar evolution, making its origin more complex. 
In fact, the abundance of fluorine is the lowest among all elements between Z = 5 and 21 and is roughly 3-4 orders of magnitude below that of C, N, O,  and Ne.
There are several plausible sources for F beyond standard stellar evolution. These include the production in the asymptotic giant branch phase (AGB) in intermediate
mass stars, production in Wolf-Rayet stars, and the production through neutrino spallation in supernovae.  The latter, known as the $\nu$-process,
is an important source for \b11, and may contribute to the abundance of \li7 as well. We combine a simple model of Galactic chemical 
evolution with a standard Galactic cosmic ray nucleosynthesis model to treat self-consistently the evolution of the Li, Be, and B isotopes.
We include massive star production of F, as well as contributions from AGB stars, and the $\nu$-process. 
Given the uncertainties in neutrino energies in supernovae, we normalize the $\nu$-process using the observed
\b11/\b10 ratio as a constraint. As a consequence, we are able to determine the relative importance
of each contribution to the F abundance.  We find that although the $\nu$-process dominates at early times (low metallicity),
the present-day F abundance is found to originate primarily from AGB stars. 
\end{abstract}


\section{Introduction}
While it is understood that virtually all of the naturally produced elements in the periodic table
are produced in stellar processes (either during or post main sequence evolution) \citep{Clayton83},
the light elements have diverse origins.  The lightest element isotopes of  D, $^3$He, $^4$He, and
$^7$Li, are produced in the early Universe during Big Bang Nucleosynthesis (BBN) \citep{2016RvMP...88a5004C,2018PhR...754....1P}. 
The slightly heavier isotopes, \li6, \li7, \be9, \b10, \b11 are produced predominantly in
interactions between cosmic rays and the interstellar medium (ISM) \citep{1970Natur.226..727R,1971A&A....15..337M,
1985ApJ...299..745W, 1990ApJ...364..568V, 2000ApJ...540..930F}, namely Galactic Cosmic Ray Nucleosynthesis (GCRN). The two forms of 
nucleosynthesis (BBN and GCRN) are related by the fact that \li7
is produced in both \citep{1992Natur.360..439O}. However, \li7 and \b11 have an additional source for production,
namely the supernova neutrino process ($\nu$-process) \citep{1990ApJ...356..272W} 
in which neutrinos produced in core collapse 
supernovae interact with C, N, O, and Ne. These elements are produced in the course of stellar evolution, but
their fragmentation by neutrino spallation can lead to appreciable amounts of \li7 and \b11. In addition, 
we expect that spallation on Ne, can produce appreciable amounts of $^{19}$F as well. 

In standard GCRN, Be and B production is expected to be secondary \citep{1990ApJ...364..568V}.  That is, 
Be, B $\propto$ O$^2$. Furthermore, if the production of Fe is related to the production of O, 
so that O/Fe is constant with respect to Fe, then Be, B $\propto$ Fe$^2$ as well.   However, 
the data tend to show a behavior which is primary in nature, Be $\propto$ Fe \citep{1988A&A...193..193R,
1990ApJ...348L..57R,1992ApJ...388..184R,1992Natur.357..379G,1993AJ....106.2309B,1996AJ....111.2106T,1997A&A...319..593M,
1999AJ....117.1549B,2000A&A...364L..42P,2000A&A...362..666P,
2001AJ....122.3115K,2009ApJ...701.1519R, 2009A&A...499..103S,
2011ApJ...743..140B}. Similarly for boron \citep{1997ApJ...488..338D,1998A&A...332.1017D,
1998ApJ...500..241G,1999A&A...343..545P}.

However, data also indicate that there is some scaling of [O/Fe] vs [Fe/H]
\citep{1998ApJ...507..805I,1999AJ....117..492B,2001ApJ...551..833I,2002A&A...390..235N}.
As a result it is in principle possible for Be and B to appear primary with respect to 
Fe, and yet secondary with respect to O \citep{1999ApJ...516..797F}.
Current data seem to favor a scaling somewhere in between pure primary
and secondary and may even indicate a change in the logarithmic slope at low
metallicity \citep{1999ApJ...516..797F,1999AJ....117.1549B,2000ApJ...540..930F,2009ApJ...701.1519R,2011ApJ...743..140B}.

As we will see below, the need for a primary component for Be and B resides
at low metallicity.  A possible source for a primary component is the infusion of low energy
nuclei (LEN)  \citep{1995Natur.373..318C,1996ApJ...468..199V,1996ApJ...456..525R,1998A&A...337..714V} or a low energy component (LEC) as it will be referred to here. 
The origins of the LEC are in superbubbles of hot metal-rich gas swept out by the collective effects of stellar winds and supernova explosions \citep{1998A&A...337..714V,
1998ApJ...509L..33H} which were originally motivated by gamma-ray observations from Orion 
\citep{1994A&A...281L...5B,1995ApJ...438L..21R,1996ApJ...462..276F,1998A&A...331..726P}.
Other primary sources for \be9 have also been considered \citep{2013PhRvL.110n1101B}. 

While the evolutionary behavior of Be and B with respect to either Fe or O is improved with the inclusion of 
a LEC, standard GCRN still suffers from two problems.
The first, the isotopic ratio of $^{11}$B/$^{10}$B, is predicted to be 2.5 \citep{1971A&A....15..337M,1975A&A....40...99M,
1985ApJ...299..745W,1988A&A...189...55A,1992ApJ...385L..13S,1993ApJ...413..562W} in contrast to the observed value of 4 \citep{1995Natur.374..337C}. 
The second problem is the overall ratio of B/Be which is generally predicted to be about 10 independent of metallicity
\citep{1993ApJ...413..562W,1995ApJ...439..854F} whereas observations place the ratio
at or above 20, though there is some some model dependence in the prediction and considerable scatter in the data.

A neat solution to both problems is an additional source of $^{11}$B \citep{1994ApJ...424..666O} from the $\nu$-process
during supernovae explosions \citep{1990ApJ...356..272W,2005PhLB..606..258H,2018ApJ...865..143S,2019arXiv190103741L,
2019ApJ...872..164K}.
Neutrino spallation on heavier elements leads to an addition source of  $^{11}$B as well as \li7 and $^{19}$F. 
There is, however, considerable uncertainty in the yields of these isotopes due to the uncertainty in the neutrino 
temperature and flux. As we will argue below, we can fix the $\nu$-process yields to avoid the overproduction of \b11.
This leaves us with definite predictions for the production of \li7 and F. 

Indeed, the production of F as it relates to GCRN is the focus of the work here. 
F is produced in both massive stars (see e.g. \citet{1995ApJS..101..181W})
and in asymptotic giant branch (AGB) stars \citep{Karakas10,Karakas14}. The $\nu$-process adds an additional
source for F \citep{1988Natur.334...45W}, and carries similar uncertainties stemming from the neutrino flux and energy. 
There have been several studies on the evolution of F in the Galaxy including the $\nu$-process \citep{2011MNRAS.414.3231K,
2011ApJ...739L..57K}. 
Recently, \citet{2018A&A...612A..16S} have studied the chemical evolution of fluorine in the solar neighborhood considering different nucleosynthetic stellar yields finding that the fluorine production is dominated by AGB stars at solar metallicity. 
Neither of these studies tied the $\nu$-process contribution to GCRN and the LiBeB elements. 

Here, we combine a simple galactic chemical evolution model with
a model of GCRN to track the evolution of the LiBeB elements together with metallicity tracers such as O and Fe.
We include the yields from AGB stars and the $\nu$-process along with the output of core collapse supernovae.
In particular, we consider the evolution of F keeping track of its production from all three sources.
We use the \b11/\b10 ratio to fix the $\nu$-process contribution which is ultimately 
small compared with the AGB contributions from intermediate mass stars.

In what follows, we first describe our model for GCRN
and chemical evolution. In section 3, we discuss our treatment of 
the evolution of the fluorine abundance and review the existing data on fluorine. 
Our results are presented in section 4.
In section 4.1, we consider the basic GCRN model with and without the LEC component and neglect the $\nu$-process. 
In sections 4.2, we show the effects of including the $\nu$-process. 
In section 4.3, we show our results on the evolution of F and we compare the relative contributions to 
the F abundance. Our conclusions are given in section 5. 

\section{Galactic Cosmic Ray Nucleosynthesis}
\label{sec:gcrn}

\subsection{Chemical Evolution}

We consider here, a simple, one-zone, closed box, chemical evolution model.
The model is described in detail in  \citet{1998ApJ...499..735L}, \citet{1999ApJ...516..797F} and also used in 
\citet{2000ApJ...540..930F}.  The mass in gas, $M_{\rm gas}$, changes due to the star formation rate,
which we take as $\psi = \nu M_{\rm gas}^n$, where we use $\nu = 0.3$ and $n = 1$, and the mass ejection rate
from deaths of stars with mass $m > 0.9 M_{\odot}$. The ejection rate is a convolution of the star formation
rate $\psi(t - \tau(m))$ and the initial mass function (IMF), $\phi(m)$ times the mass ejected as a function of the initial stellar mass, $m$. 
The lifetime of the star $\tau(m)$ is taken into account to determine the ejection rate. 
We use a simple power law IMF, $\phi \propto m^{-2.7}$ which is normalized by integrating $m \phi$ over the mass
range $0.1 - 100 M_\odot$. While there is a slight difference in the Be and B abundances when we consider
an open box model with outflows \citep{1999ApJ...516..797F}, that will not be our focus here.

The element isotopes we consider here are H, He, C, N, O, and Fe, as well as Li, Be, B, and F
which are affected by cosmic ray nucleosynthesis (in the case of Li, Be, and B) and the 
$\nu$-process (in the case of Li, B, and F). We use the yields tabulated in \citet{1995ApJS..101..181W}
for the high mass stars ($M > 8 M_\odot$) produced by Type II supernovae. We use the yields from \citet{Karakas10,Karakas14}
for intermediate mass stars ($1 - 8 M_\odot$) which are particularly important for the evolution of F. 
The importance of intermediate stellar mass yields was recently demonstrated for the production of the heavier
Mg isotopes \citep{2019MNRAS.484.3561V}. 
Finally, we use the recent set of yields from \citet{2018ApJ...865..143S} for the $\nu$-process yields of \li7, \b11, and F. 
Due to the sensitivity to the high energy tail of the neutrino energy distributions, these authors
provide two sets of yields denoted as low and high energy which depend on their choice of neutrino temperatures.
In this work, we consider both sets of yields. We also considered the yields in \citet{1995ApJS..101..181W} 
for the $\nu$-process. We find that these result in abundances which are very similar to the abundances found using
the high energy yields of \citet{2018ApJ...865..143S}. 

\begin{figure}
\begin{center}
\epsfig{file=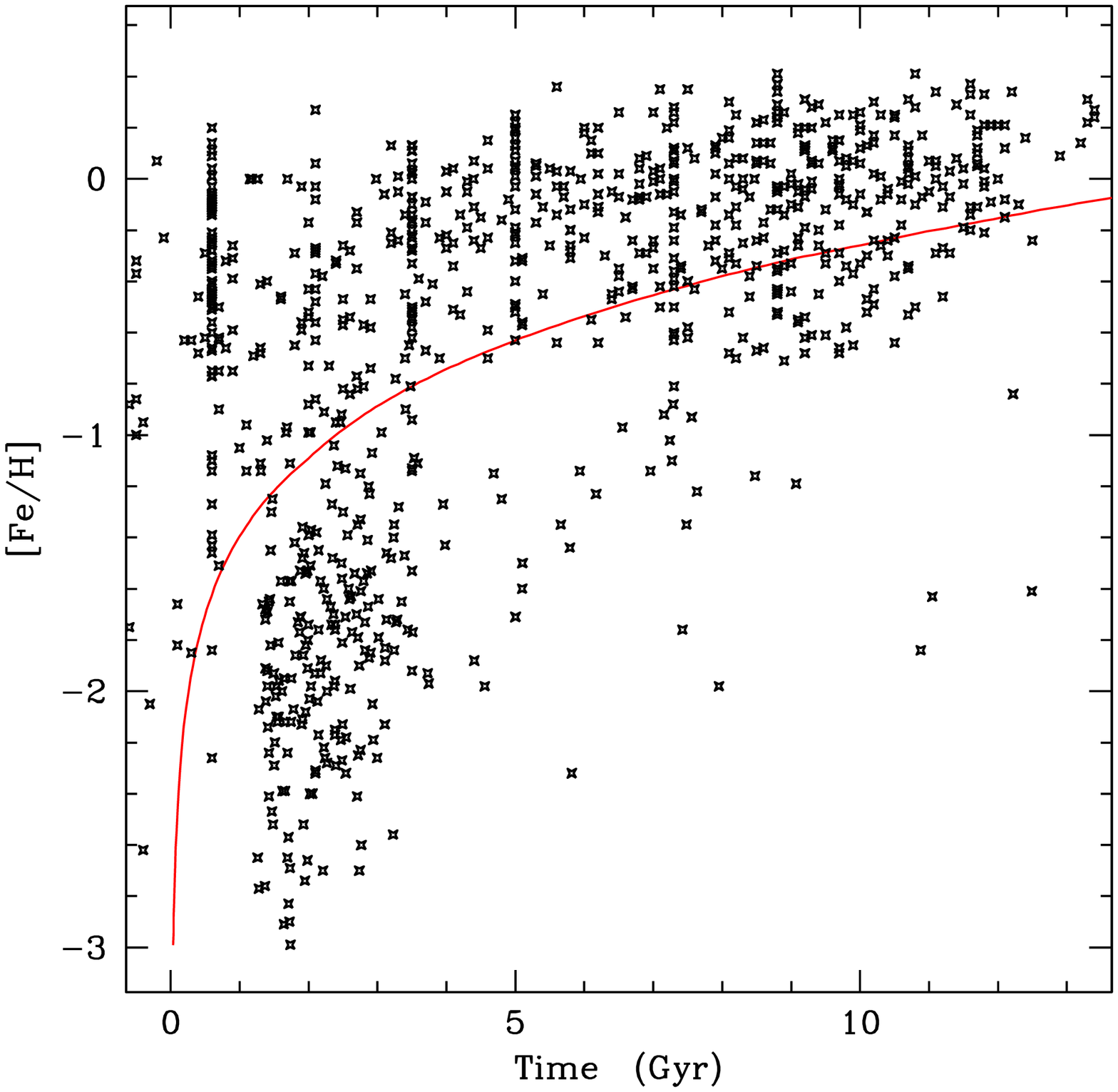, height=3in}\\
\epsfig{file=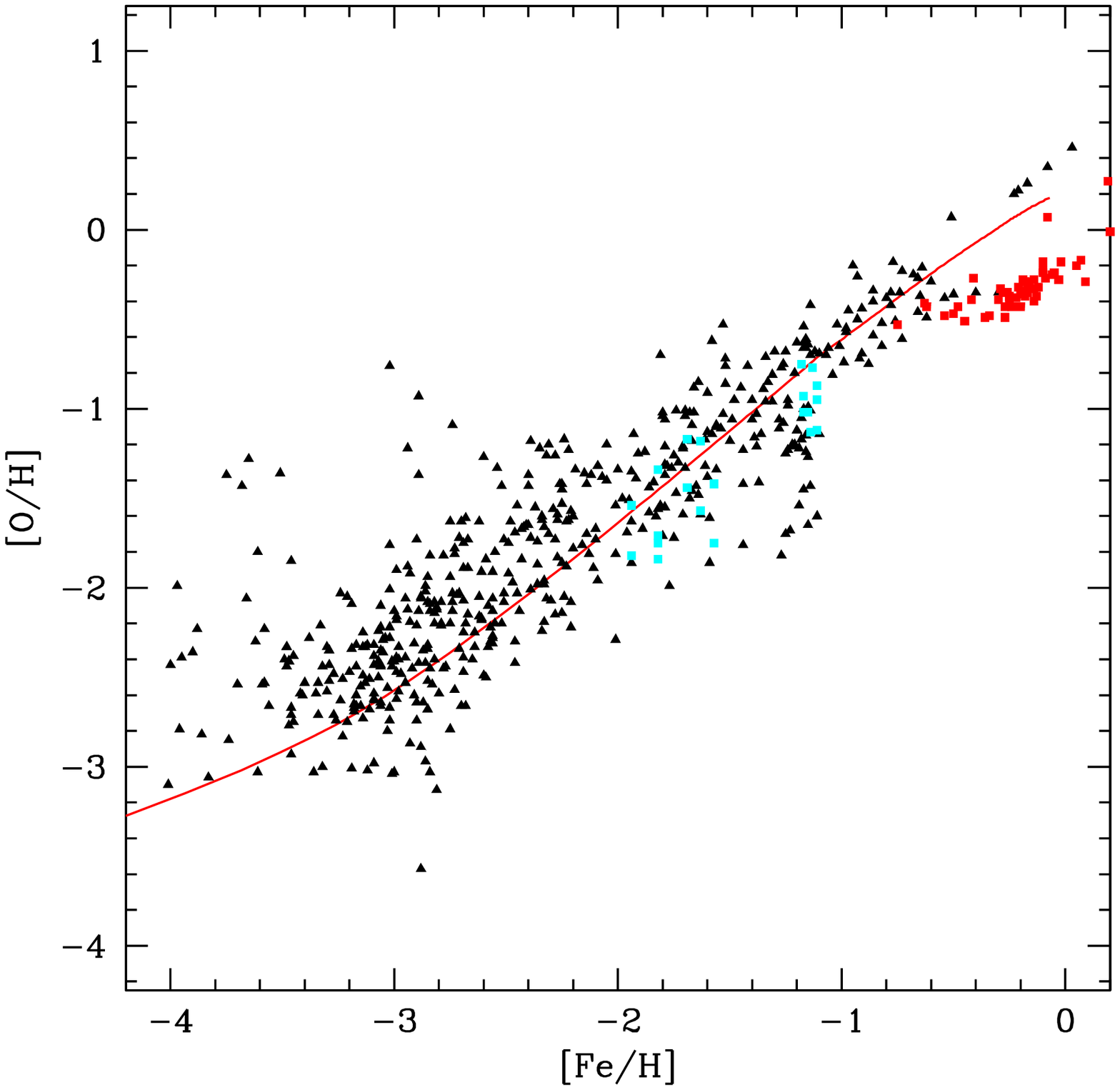, height=3in}
\end{center}
\caption{The abundances of [Fe/H] vs time and [O/H] as a function of [Fe/H] in a simple model 
for Galactic chemical evolution. Data on O/H are taken from \citet{2007MNRAS.380L..40F,2010ApJ...721....1P,2011MNRAS.417.1534C} 
and data on Fe/H from \citet{raf12b,2014A&A...562A..71B}. We have also included here O vs Fe data
taken from fluorine studies shown by the red squares \citep{2017ApJ...835...50J} and cyan squares \citep{2013ApJ...763...22D, 2019ApJ...876...43G}.
Solar values for all abundances are taken from \citet{1989GeCoA..53..197A}.}
\label{fig:OFe}
\end{figure}

The top panel of Fig. \ref{fig:OFe} shows the resulting evolution of [Fe/H] with respect to time. 
The data are taken from \citet{raf12b,2014A&A...562A..71B}. The lower panel shows the evolution of [O/H]
with respect to [Fe/H]. The oxygen data are taken from \citet{2007MNRAS.380L..40F,2010ApJ...721....1P,2011MNRAS.417.1534C}.
As one can see, both O and Fe are reasonably well fit in this simple model. Note that in the lower panel,
we have also included O and Fe data taken from the fluorine studies discussed below.
We only remark here, that  the oxygen abundance in these data are systematically low
with respect to other data and to our evolutionary curve. This is particularly true at higher
metallicity.  Though we do not show it here, 
the present day gas mass fraction is 6\%. 

\subsection{GCRN}

Despite the many uncertainties that go into the origin and propagation of cosmic rays,
the results of cosmic ray nucleosynthesis are relatively robust. 
Here, we use the formalism outlined in  \citet{1994ApJ...435..185F,1998ApJ...499..735L,2000ApJ...540..930F}.

The production rate of a particular isotope per unit volume is
\beq
\frac{d}{dt} n_\ell = \sum_{ij} n_i 
  \int \ dE \  \sigma_{ij}^{\ell}(E) \ \phi_j(E) \ S_\ell (E) \, ,
\eeq
where $n_i$ is the ISM number density of species $i$,
$\phi_j(E)$  is the cosmic ray flux for particle
species $j$,
and $\sigma_{ij}^{\ell}(E)$ is the cross section for
the spallation reaction $i + j \rightarrow \ell + \cdots$
 \citep{1984ADNDT..31..359R}.
The factor $S_\ell(E)$
is the probability that the daughter nucleus $\ell$ is stopped
and thermalized in the ISM before it can escape from the Galaxy.
The cosmic ray flux is determined from an initial source spectrum which is proportional to the
star formation rate. The source spectrum is taken as a power law in energy \citep{1997ApJ...487..197E}
and propagated in a leaky box model.
Important processes include cosmic ray $p, \alpha$ on C,N,O in the interstellar medium (ISM)
as well as reverse processes of cosmic ray C,N,O on ISM $p$ and $\alpha$, though in standard GCRN,
these are subdominant. Also of importance for the production of Li is $\alpha$ + $\alpha$. 

As noted earlier, an additional contribution to the production of the LiBeB elements can come 
from the acceleration of particles within superbubbles formed by the collective explosions of supernovae. 
While particle production within the superbubble is rare due to low density in these regions,
the synthesis of LiBeB occurs as highly enriched supernova ejecta escape the superbubble and traverse the 
more dense outer shells. We assume that the composition of this ejected material is the same as that 
of a $40 M_\odot$ star and use the yields in \citet{1995ApJS..101..181W}. This is a primary source for Be and B. 

Finally, we include the $\nu$-process in our calculation of the yields for Li and B.  
The neutrino process \citep{1990ApJ...356..272W}
provides a primary source of production for \li7 and \b11 in supernovae.
As it streams from the supernova core, a flux of neutrinos
traverses the outer shells of material.  
As neutrinos pass through the carbon shell,
they collide inelastically with the
\iso{C}{12} nuclei, and 
by removing a nucleon: $\iso{C}{12} + \nu \rightarrow \b11 + p + \nu^\prime$ make primary \b11.
Removing a neutron produces $\iso{C}{11}$ which $\beta$ decays to \b11. 
Some \b10 and \be9 are also produced but little survives the final explosive process. 
Thus the $\nu$-process is an efficient mechanism for altering both the \b11/\b10 and B/Be ratios. 

Some \li7 is also produced in the He shell. He is spalled by neutrinos to make $^3$He and $^3$H which is 
subsequently burned to \li7 and \be7 (which decays to \li7) as well as some production of \b11 and $\iso{C}{11}$. 
As we will see, however, the amount of \li7 produced is relatively minor if the production of \b11 is limited by the 
\b11/\b10 ratio. F is predominantly produced in the Ne shell by neutrino spallation making both $^{19}$F and $^{19}$Ne which
decays to $^{19}$F. 

We use both the low and high energy yields of \citet{2018ApJ...865..143S}. 
The latter produce results similar to that using the \citet{1995ApJS..101..181W} yields. 
We view the difference in these yields as an inherent uncertainty in the $\nu$-process. 
As a result we interpolate between the two sets of yields to obtain the correct \b11/\b10 ratio.

Each of our three components used in the production of the LiBeB elements carries some uncertainty.
As in \citet{2000ApJ...540..930F}, we fix the normalizations of each component as follows.
As we noted earlier, the overall cosmic ray flux is proportional to the star formation rate,
and while the shape of the spectrum is fixed its overall normalization is adjusted so that the
Be abundance today is matched to the observed value. 
Our second normalization concerns the LEC. The relative amount for primary production
is also uncertain, and we add this component so that at the slope of [Be/H] vs [Fe/H] matches the observed
slope (which lies between purely primary and secondary). Finally,
we interpolate between the low (L) and high (H) $\nu$-process yields
so that the adopted yield is $\alpha$ L + (1 - $\alpha$) H. Thus $\alpha = 1$ corresponds to 
the low energy yields while $\alpha = 0$ corresponds to the high energy yields. The same interpolation is
made for all $\nu$-process products.

\section{Fluorine}
\label{F}

\subsection{Sources for Fluorine}
\label{sources}

Of the elements produced in stars ($A \ge 5$), the abundance of 
fluorine is anomalously low. This is mainly due to the fact that fluorine is easily
destroyed through either $p$ or $\alpha$ captures. 
As hinted above, there are (at least) three sources of elemental fluorine. 
Fluorine is produced in massive stars in the helium burning shell \citep{2000A&A...355..176M}.
While the yields of F in stars with masses $M < 20 M_\odot$ are relatively low,
more massive stars which become Wolf-Rayet stars can eject F through stellar winds.
The production of $^{19}$F proceeds through one of several nuclear chains.
For example, 
\beq
{}^{14}{\rm N} (\alpha, \gamma) {}^{18}{\rm F}(e^+, \nu) {}^{18}{\rm O}(p,\alpha) {}^{15}{\rm N}(\alpha,\gamma) {}^{19}{\rm F} \, .
\eeq
Although $^{19}$F is easily destroyed through $^{19}$F$(\alpha, p) {}^{22}$Ne,
the strong winds in massive stars allow fluorine to be ejected before its destruction.

The second and most dominant source of fluorine is the production during
thermal pulses on the AGB \citep{1992A&A...261..157F,1996A&A...311..803M}.
In this case, the fluorine survives so long as the destruction timescale is
shorter than the pulse time. This implies that $^{19}$F production is most efficient
in the early AGB stages before the temperatures exceed $\sim 3 \times 10^8$K. 
Here we use the yields from \citet{Karakas10,Karakas14} for intermediate mass stars. 

The third possible source for F is the neutrino process \citep{1988Natur.334...45W}. 
Neutrinos emerging from core collapse, inelastically scatter with nuclei
(of interest here is $^{20}$Ne) producing an excited state which
subsequently decays to either $^{19}$F or $^{19}$Ne through
$p$ or $n$ emission respectively. Much of the $^{19}$Ne is destroyed by 
subsequent nuclear reactions. The most significant uncertainty is the temperature of 
the neutrinos. Additional destruction of both  $^{19}$F and $^{19}$Ne
occurs during the passage of the shock.  
Since AGB stars provide both primary and secondary sources of F, it is difficult to use 
observations to distinguish between the AGB and SN sources for F (which are primary). 
For the supernova yields of F, we use
\citet{2018ApJ...865..143S} who provide yields for F (as well as \li7 and \b11)
neglecting the neutrino process, and for two sets of neutrino temperatures.
We perform calculations using all three sets of yields.

\subsection{Fluorine Data}
\label{data}

The first observations of fluorine outside the solar system were made using the infra-red
vibration-rotation line of the molecule HF in several K stars and AGB stars \citep{1992A&A...261..164J}.
The data from these observations 
show a correlation between F/O and C/O and confirmed AGB stars as (at least) one source of fluorine production. 
These observations also support the correlation between F and s-process production \citep{1998A&A...334..153M}. 
While observations reported in \citet{2011ApJ...737L...8A,2012A&A...538A.117R,2013ApJ...763...22D} 
show a correlation with the s-process,
it is argued that the dependence on metallicity differs from the predictions of low metallicity AGB models.

Measurements of the fluorine abundance outside the Galaxy and in the globular cluster $\omega$ Centauri
were made in \citet{2003AJ....126.1305C}. On the basis of a large scatter of [F/H] with respect to [Fe/H]
and depleted values of [F/H[ in $\omega$ Centauri, these authors argued against AGBs as a dominant source 
of F.  Instead, they argued that the trend in the data agree reasonably well with chemical evolution models including 
the $\nu$-process \citep{1995ApJS...98..617T,2001A&A...370.1103A}, though the present day
abundance falls short of solar. 
\citet{2013ApJ...765...51L} provide data from low metallicity giants and argue for some 
$\nu$-process contributions at least at low metallicity. 
However, when the data was compared with
more recent chemical evolution models which incorporate both massive and intermediate mass stars, 
the AGB contribution was found to be essential \citep{2004MNRAS.354..575R,2005NuPhA.758..324R}. 
We note that some earlier data on F abundances are systematically off due to the use of inconsistent
molecular data \citep{2019ApJ...876...43G}. 

Measurements of fluorine in the ISM along the line of sight of a region affected by 
past supernovae showed no sign of enhanced F \citep{2005ApJ...619..884F,2007ApJ...655..285S}, also disfavoring 
the $\nu$-process as a dominant source. In contrast, \citet{2005A&A...433..641W} found enhanced fluorine abundances in 
hot post-AGB stars supporting evidence for AGB production. 
More recent data in C-rich low metallicity stars \citep{2011ApJ...729...40L}, carbon stars at low and solar metallicity
\citep{2015A&A...581A..88A}, and in extragalactic AGB stars \citep{2011ApJ...737L...8A}
also point to AGB production as do observations in the solar neighborhood \citep{2014ApJ...789L..41J,2017ApJ...835...50J}.
Data from the globular cluster M22 \citep{2012A&A...540A...3A,2013ApJ...763...22D} show considerable dispersion and fluorine there is
also ascribed to AGB production. Recently, data from the globular cluster M4 is given in \citet{2019ApJ...876...43G}.

Finally, a number of recent observations of fluorine point to sources other than AGB stars.
These include open clusters \citep{2013AJ....146..153N}, the thin disk \citep{2015AJ....150...66P}, and the bulge
\citep{2014A&A...564A.122J}, where the latter point specifically to Wolf-Rayet contributions.
We note that these are relatively high metallicity systems. 

As one can see, despite a growing body of data, a complete understanding of the origin 
of fluorine remains lacking. Though, there is most certainly a strong contribution to present-day
fluorine from AGB nucleosynthesis, the role of the $\nu$-process remains uncertain.
In what follows, we will tie the strength of the $\nu$-process contribution to F 
 to the GCRN production of \b11.

\section{Results}

\subsection{Standard Cosmic Ray Nucleosynthesis}

\begin{figure}
\begin{center}
\epsfig{file=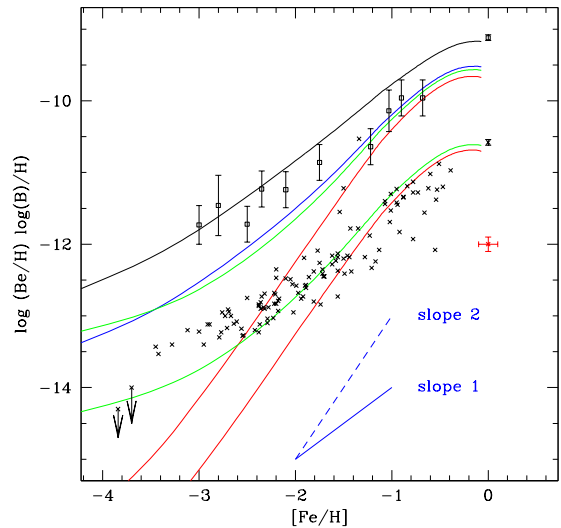, height=3in}
\end{center}
\caption{The abundances of B/H and Be/H as a function of [Fe/H] in GCRN.
The red curves represent the evolution in standard GCRN. The green curves include the effects of a LEC.
The blue and black curves for B include the effects of $\nu$-process production for the low and high
yields given in \citet{2018ApJ...865..143S}. Sources for the data are given in the text.}
\label{fig:BeBFe}
\end{figure}

We start by considering the production of Be and B in standard GCRN
embedded in a simple model of chemical evolution as described above.
The evolution Be/H and B/H with respect to [Fe/H] is shown in Fig. \ref{fig:BeBFe}.
The two red curves show the evolution of B (upper) and Be (lower) ignoring any LEC
or $\nu$-process contributions. The B data are the upper set of points (with error bars)
taken from \citet{1992ApJ...401..584D,1997ApJ...488..338D,1998ApJ...500..241G},
and the Be data correspond to the lower set of points taken from 
\citet{2011ApJ...743..140B,2009ApJ...698L..37I,2019A&A...624A..44S}.
The typical error bar for the Be data is shown by the cross at (0,-12). 
For ease of comparison, the blue line segments show the slope for 
primary (1) and secondary (2) evolutions. As expected, in standard GCRN,
both B and Be evolve as secondary elements, and while the evolutionary curves match the 
data reasonably well at higher metallicities ([Fe/H] $> - 1.5$), they fall well short at low metallicities. 
Solar data is taken from \citet{1989GeCoA..53..197A}.

\begin{figure}
\begin{center}
\epsfig{file=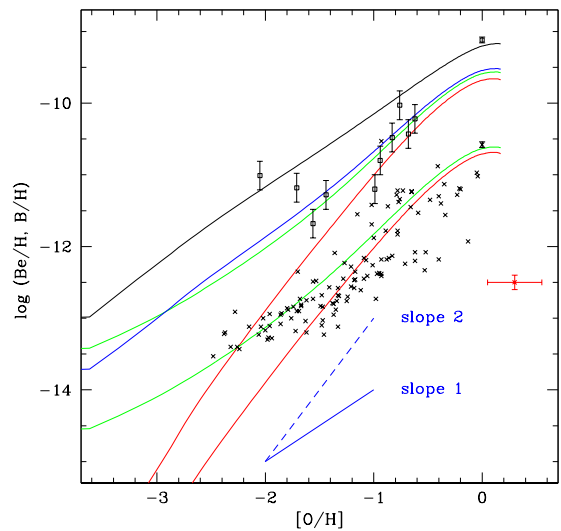, height=3in}
\end{center}
\caption{As in Fig. \ref{fig:BeBFe}, for the abundances of B/H and Be/H as a function of [O/H].}
\label{fig:BEBO}
\end{figure}

As first argued in \citet{1999ApJ...516..797F}, the correlation of Be and B is more naturally matched to 
oxygen than it is to [Fe/H]. In Fig. \ref{fig:BEBO}, we show the evolution of Be/H  and B/H as a function of [O/H].
While the computed slope is still 2 (secondary production), the data for Be/H fall off more steeply
with decreasing [O/H] as the [O/Fe] is not constant at low [Fe/H].  The curves only fall significantly 
below the data for [O/H] $< -2$.  Thus there appears to be evidence for some primary component
at low metallicity \citep{1995Natur.373..318C, 1998A&A...337..714V, 2000ApJ...540..930F,2000A&A...356L..66P,2000A&A...362..786P,2001AJ....122.3115K}.
The green curves in both Figs. \ref{fig:BeBFe} and \ref{fig:BEBO} include the LEC component
which is effective primarily at low metallicity. 
Note that the overall cosmic ray flux has been normalized so that Be/H today match the observed data point
at [Fe/H] = 0 when the LEC is included. The strength of the LEC component is normalized to produce the correct slope for 
Be/H at low metallicity.

\begin{figure}
\begin{center}
\epsfig{file=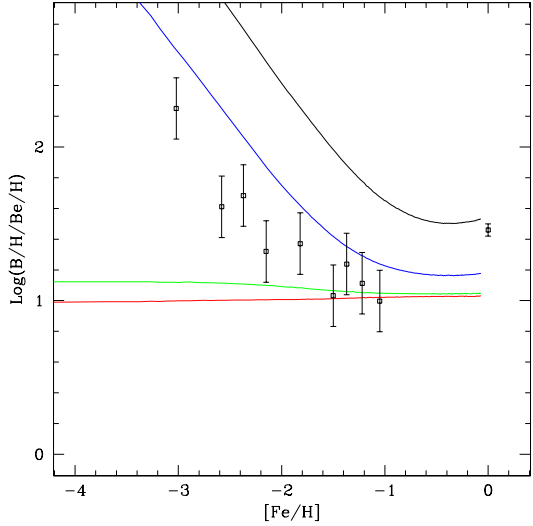, height=3in}
\end{center}
\caption{As in Fig. \ref{fig:BeBFe} for the evolution of B/Be as a function of [Fe/H].}
\label{fig:B/BeFe}
\end{figure}

When one compares the ratios of GCRN-produced elements and isotopes, it is clear
that standard GCRN even with a LEC is insufficient for fully explaining the
abundances of B. For example, consider the evolution of B/Be vs [Fe/H] as shown in
Fig. \ref{fig:B/BeFe}. As in Fig. \ref{fig:BeBFe}, the red curve corresponds to standard GCRN
which predicts a B/Be ratio of about 10, independent of metallicity, and the green curves includes the LEC contribution.
Here, we see again that the LEC is important at low metallicity. 
The data however, show B/Be ratios far in excess of 10. At low metallicity,
some examples of B/Be $\sim 100$ are known. Thus it would appear that a source for B
other GCRN is needed. The B/Be values correspond to the IRFM1 data described in detail in \citet{2000ApJ...540..930F}. 

\begin{figure}
\begin{center}
\epsfig{file=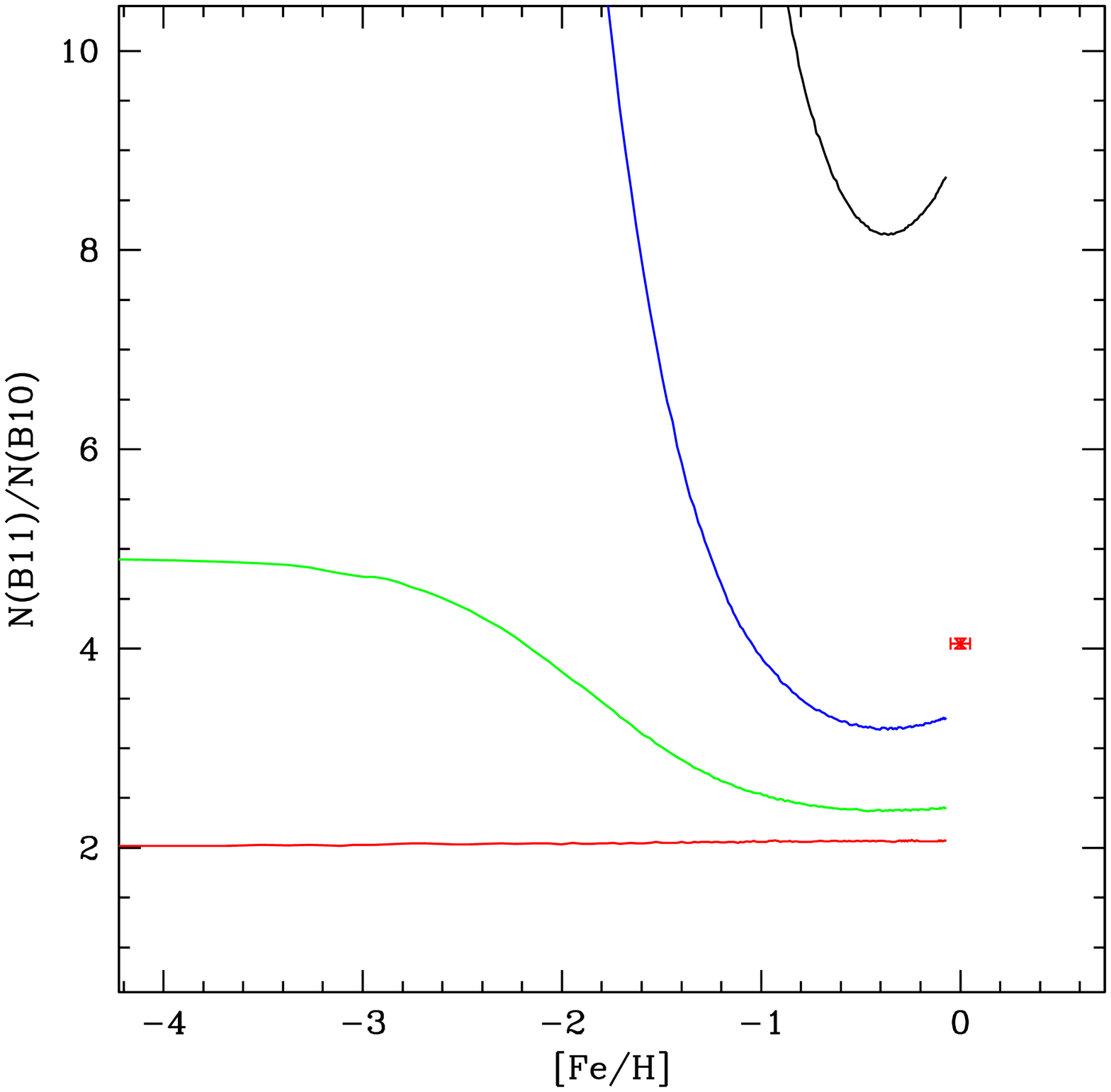, height=3in}
\end{center}
\caption{As in Fig. \ref{fig:BeBFe} for the evolution of \b11/\b10 as a function of [Fe/H].}
\label{fig:1110}
\end{figure}

A similar conclusion is drawn when we compare the isotopic ratio of 
\b11/\b10 as shown in Fig. \ref{fig:1110}. Once again, the red curves show
the evolution of the 11/10 ratio in standard GCRN and the green curve includes the LEC.
GCRN predicts \b11/\b10 = 2. While a LEC can amplify this ratio at low metallicity,
it only marginally increases the solar ratio to roughly 2.5, whereas the observed solar 
value is $\b11/\b10 = 4.05 \pm 0.16$ \citep{1995Natur.374..337C}. 
Thus the data call specifically for additional source of $\b11$ with respect to standard GCRN.

\begin{figure}
\begin{center}
\epsfig{file=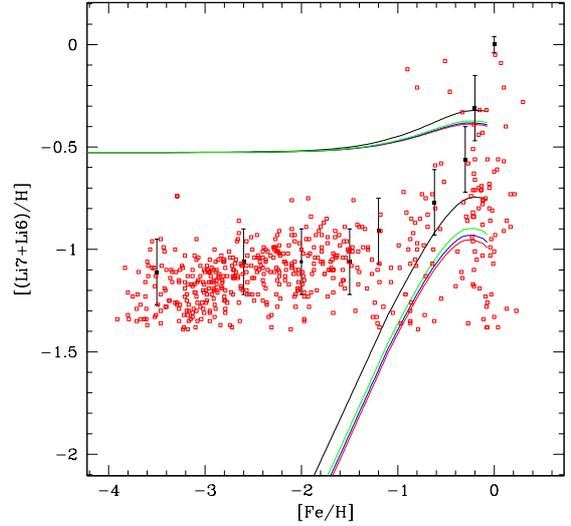, height=3in}
\end{center}
\caption{As in Fig. \ref{fig:BeBFe} for the evolution of [Li/H] as a function of [Fe/H]. 
Red points come from the SAGA database compilation \citep{2008PASJ...60.1159S}. Black points with error bars are representative of the mean observations. }
\label{fig:Li}
\end{figure}

For completeness we show the evolution of \li7 with respect to [Fe/H] in Fig. \ref{fig:Li}.
The LEC (green curve) adds very little Li to the standard GCRN production (red).
Both of these ignore the primordial Li abundance which is added to the GCRN (and $\nu$-process production)
contributions in the nearly horizontal curves coming from BBN \citep{2016RvMP...88a5004C,2018PhR...754....1P}. 
The primordial lithium problem is clearly seen as the BBN abundance lies significantly above the data \citep{2008JCAP...11..012C}.
Furthermore, the early evolution of Li is swamped by the primordial abundance leaving the plateau in tact. 
Note that even with the high $\nu$-process yields (black curves), the solar abundance is not achieved implying the need for
additional late time sources of Li. 

\subsection{$\nu$-Process}

From the above discussion, the addition of the $\nu$-process production of \b11
can clearly remedy both problems relating to the B/Be and \b11/\b10 ratios. 
As we indicated earlier, we use the $\nu$-process yields given in \citet{2018ApJ...865..143S}. 
Due to the sensitivity to and uncertainty in the neutrino temperatures, \citet{2018ApJ...865..143S}
provide two sets of yields. Their high energy yields assume
the following neutrino temperatures: $T_{\nu_e} = 4$ MeV, $T_{{\bar \nu_e}} = 5$ MeV and $T_{\nu_{\mu,\tau}} = 6$ MeV.
These give results similar to those found in \citet{1990ApJ...356..272W} and \citet{2005PhLB..606..258H}. 
The low energy yields assume  $T_{\nu_e} = 2.8$ MeV, and $T_{{\bar \nu_e},{\nu_{\mu,\tau}}} = 4$ MeV.

Our results including $\nu$-process production of \b11 for B/H vs [Fe/H] are shown 
in Fig. \ref{fig:BeBFe}. The blue (black) curves correspond to the low (high) energy yields given
in \citet{2018ApJ...865..143S}\footnote{Note that the blue and black curves omit the LEC contribution,
and show only the standard GCRN contribution augmented by the $\nu$-process. This is the case in 
all figures showing the effect of the high and low $\nu$-process yields.}. As one can see, the low energy yields add little to the total 
B/H abundance whereas the high energy yields increase the B/H abundance by as much as a factor of 5.
This effect is also seen in Fig. \ref{fig:BEBO} for the evolution of B/H with respect to [O/H].
Similarly, for \li7, the high energy yields increase the late-time abundance by a factor $\lesssim 2$
as seen in Fig. \ref{fig:Li}. When the BBN abundances are included, the effect is greatly diluted,
and the increase in the \li7 abundance due to the $\nu$-process is of order 15\%. 

More striking is the effect on the element ratios. In Figs. \ref{fig:B/BeFe} and \ref{fig:1110},
we show the B/Be and \b11/\b10 ratios as functions of [Fe/H] for the high (black) and low (blue) energy
yields. As one can see, the B/Be ratio is significantly increased, and ratios higher
than 20 are expected at low metallicity. Since the $\nu$-process produces very little Be (and we have not included
this contribution), boron (in the form of \b11) is expelled into the ISM very early (due to the first 
generations of type II supernovae) with very little Be present. The same is true for the ratio 
\b11/\b10.

 One of our key motivations for including the $\nu$-process is deficiency in the
 \b11/\b10 ratio.
 Including the $\nu$-process, we find that the present-day boron isotopic ratio is 
 8.7 for the high energy yields and 3.4 for the low energy yields. 
 We normalize the ``neutrino temperature" and interpolate between the two 
 sets of yields. To that end, we assume that the true yield is given by
 \beq
 Y = \alpha Y_{\rm low} +  (1-\alpha) Y_{\rm high} \,
 \eeq
so that $\alpha = 1 (0)$ corresponds to using the low (high) energy yields. 
We fix the value of $\alpha$ by demanding the correct \b11/\b10 ratio today.
We find $\alpha = 0.9$ to obtain \b11/\b10 = 4. 

\begin{figure}
\begin{center}
\epsfig{file=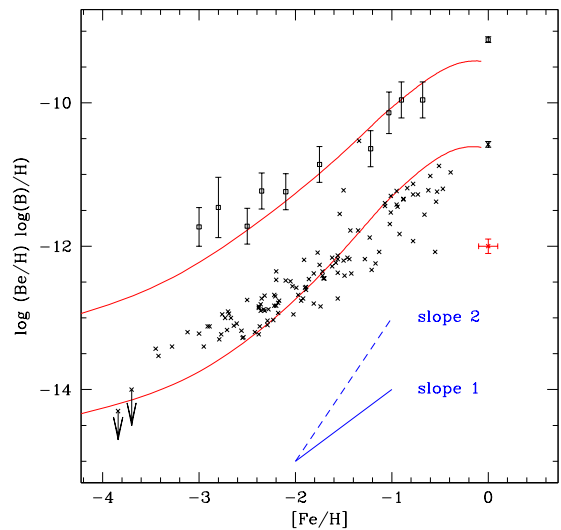, height=3in}
\end{center}
\caption{The abundances of B/H and Be/H as a function of [Fe/H] in standard GCRN including the LEC and
$\nu$-process production of \b11. 
The $\nu$-process contribution has been scaled to yield the correct \b11/\b10 ratio.}
\label{fig:BeBFeb}
\end{figure}

\begin{figure}
\begin{center}
\epsfig{file=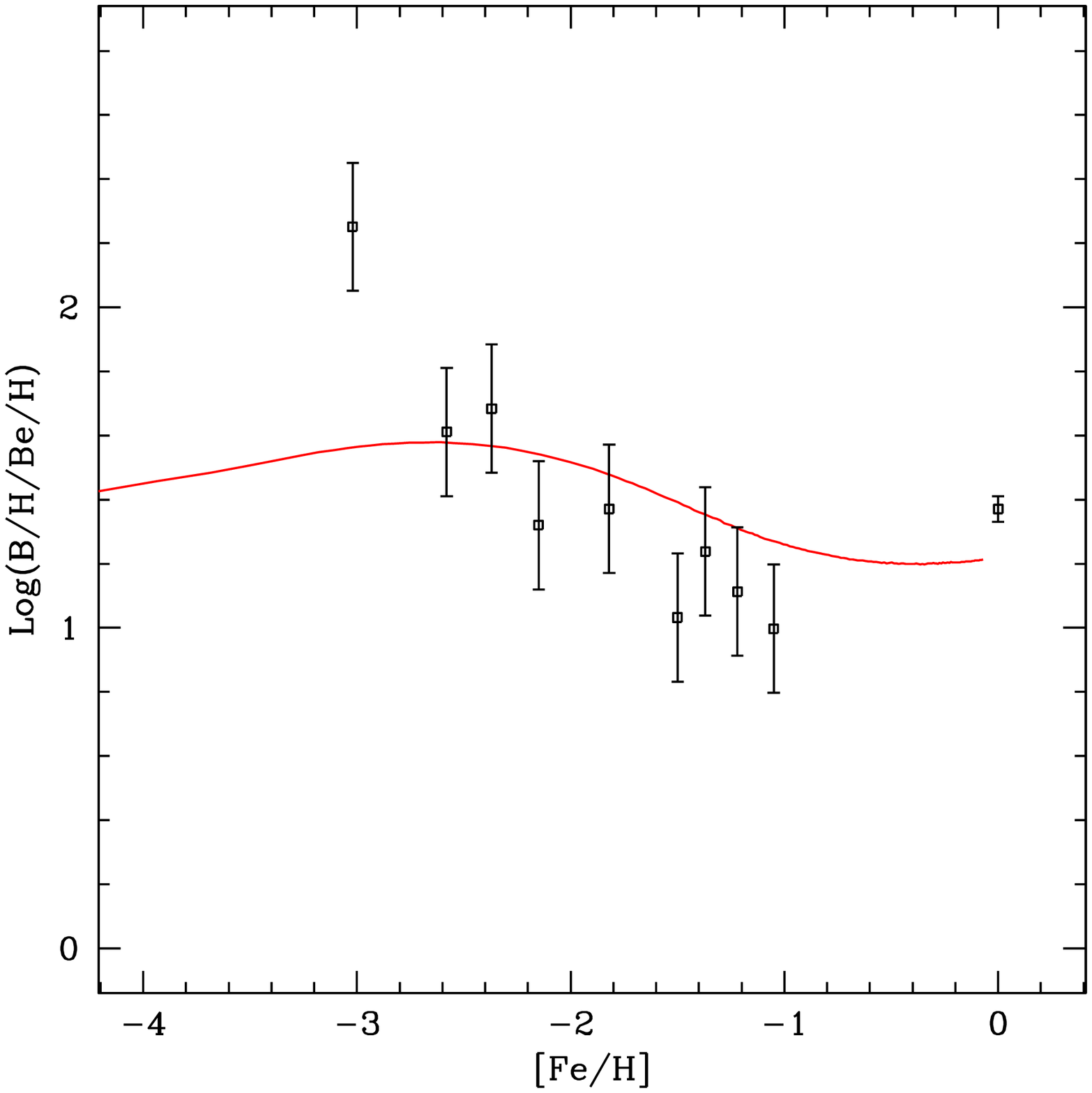, height=3in}
\end{center}
\caption{As in Fig. \ref{fig:BeBFeb} for the evolution of B/Be as a function of [Fe/H].}
\label{fig:B/Beb}
\end{figure}

\begin{figure}
\begin{center}
\epsfig{file=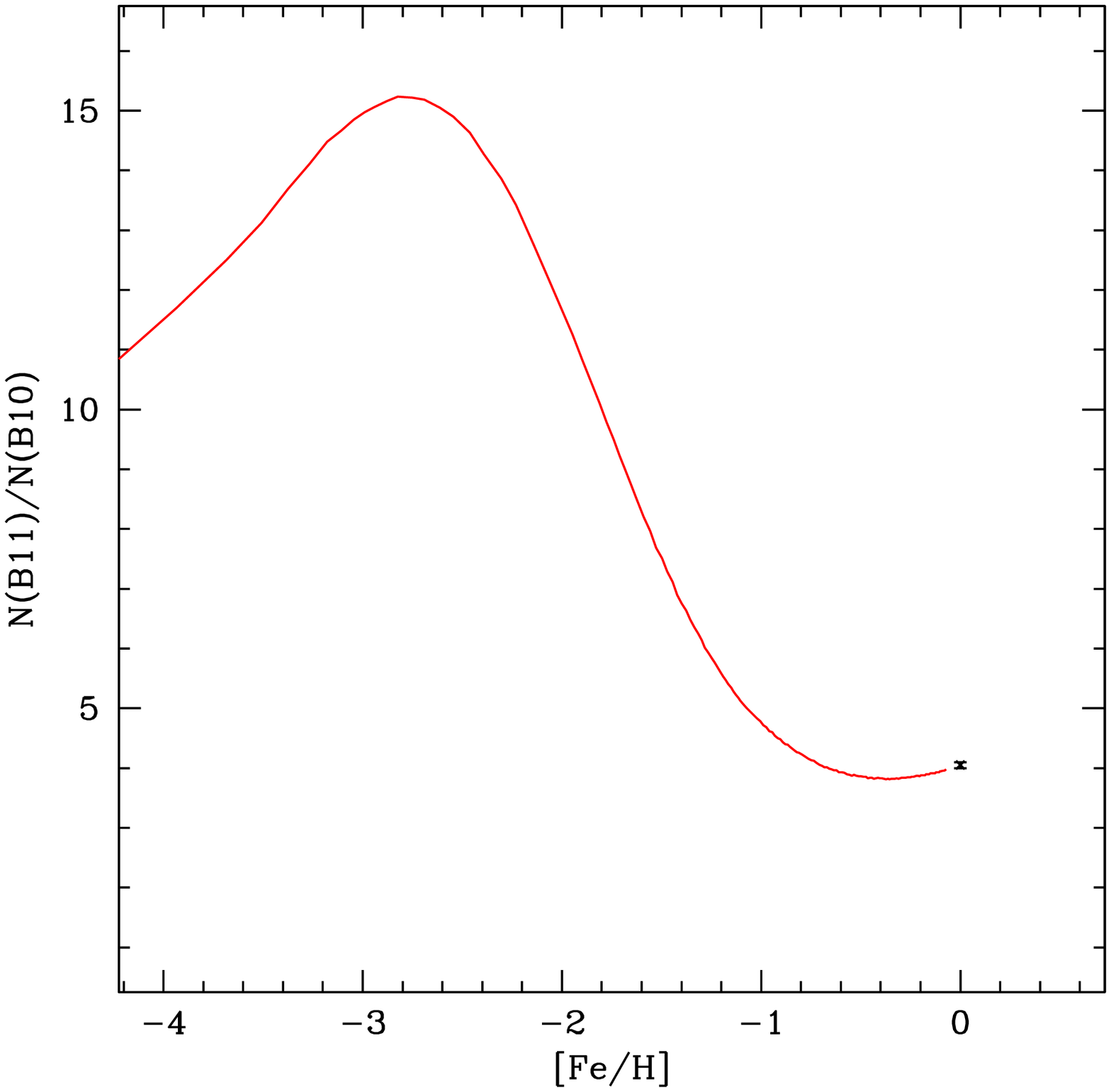, height=3in}
\end{center}
\caption{As in Fig. \ref{fig:BeBFeb} for the evolution of \b11/\b10 as a function of [Fe/H].}
\label{fig:1110b}
\end{figure}

Our final results for the evolution of Be and B are shown in Figs. 
\ref{fig:BeBFeb}, \ref{fig:B/Beb}, and \ref{fig:1110b}. The single curve in each
figure shows the evolution including standard GCRN normalized by the present 
Be abundance, an LEC component to match the slope of Be vs Fe at low metallicity,
and the $\nu$-process with $\alpha = 0.9$ to obtain the correct \b11/\b10 ratio.
As seen in Fig.  \ref{fig:BeBFeb}, the evolution of B and Be are in reasonably 
good agreement with the data. 


The B/Be ratio shown in Fig. 
\ref{fig:B/Beb} is also in good agreement with the data, though there are some objects with
anomalously high B/Be ratios. Finally, we see the \b11/\b10 ratio in Fig. 
\ref{fig:1110b} where the final ratio at [Fe/H] = 0 is used as an input to derive $\alpha = 0.9$ and
hence adjust all of the $\nu$-process yields.

\subsection{Fluorine}

\begin{figure}
\begin{center}
\epsfig{file=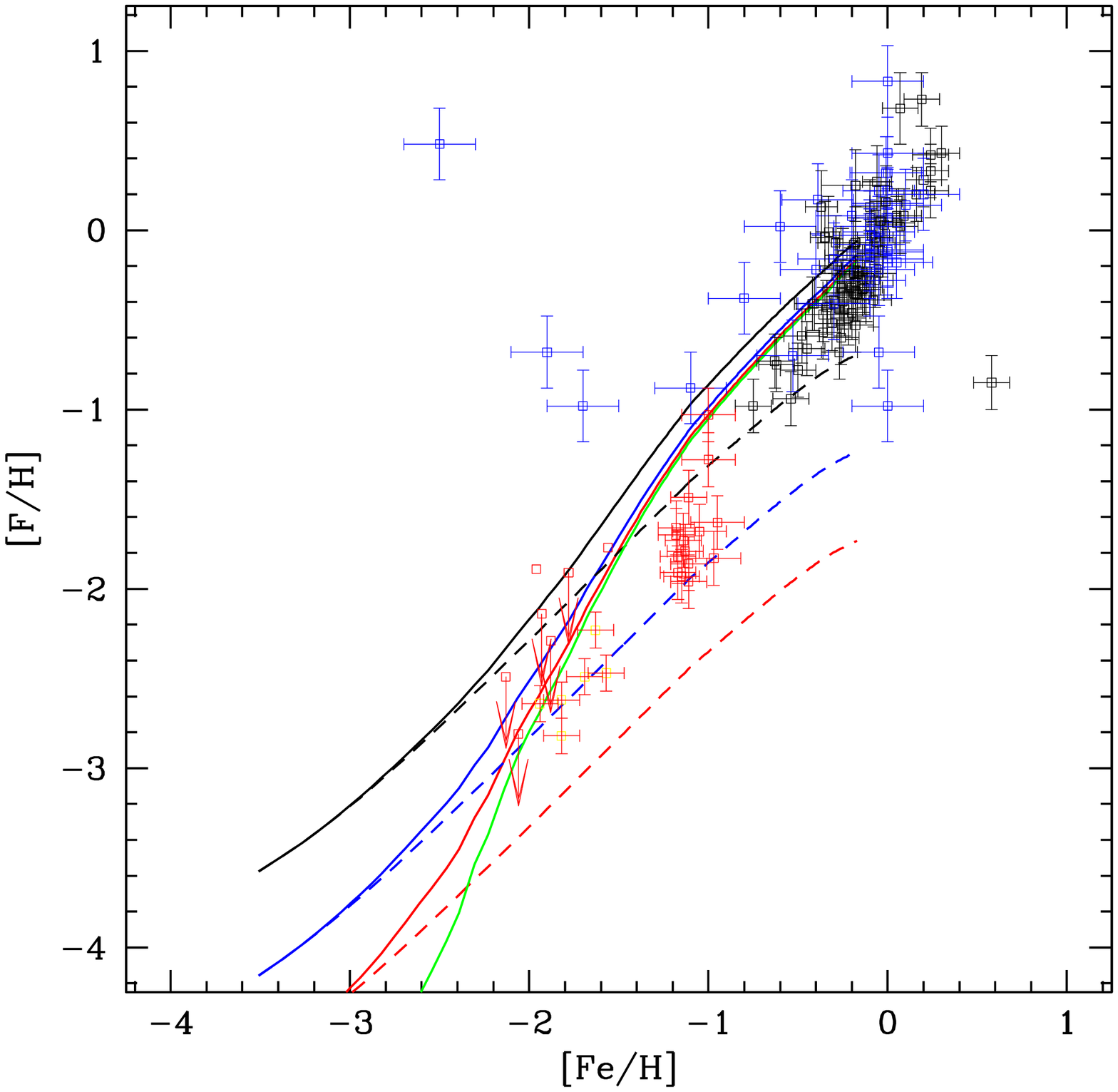, height=3in} \\
\epsfig{file=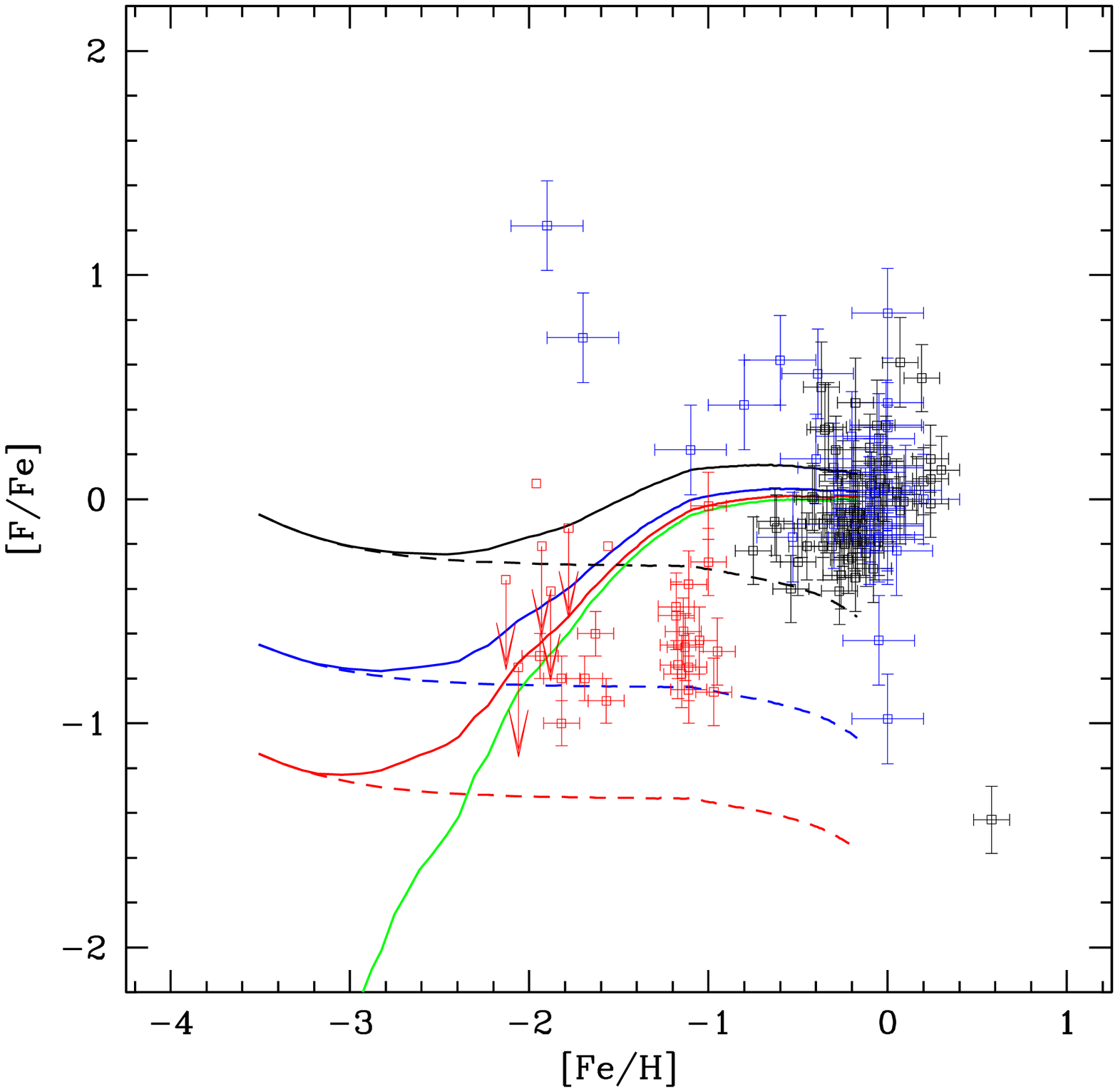, height=3in}
\end{center}
\caption{The evolution of [F/H] as a function of [Fe/H] (upper panel)
and the evolution of [F/Fe] vs [Fe/H] (lower panel).
The green curves omit the $\nu$-process contributions.
The blue and black curves include the effects of $\nu$-process production for the low and high
yields given in \citet{2018ApJ...865..143S}. The thin green line shows only the AGB contribution, while 
the dashed curves omit the contributions from AGB stars.
Sources for the data are given in the text. }
\label{fig:FFea}
\end{figure}

Finally, we present our results for the evolution of fluorine. 
In the upper panel of  Fig. \ref{fig:FFea}, we show the evolution of [F/H] vs [Fe/H].
The dashed curves show the evolution in the absence of any AGB production of F. 
The green, blue and black curves correspond to no $\nu$-process, low energy, and high energy
yields respectively. Taken alone, the $\nu$-process contributes substantially to the F abundance at low metallicity. 
The lower panel is similar and shows the evolution of [F/Fe] vs [Fe/H].
While the neutrino-process may be capable of explaining some data at low metallicity, 
even the high energy yields fall far short of the abundances at (and above) solar metallicity.
The F data are taken from i) \citet{ 2011ApJ...737L...8A, 2015A&A...581A..88A}, blue points, ii)
 \citet{  2012A&A...538A.117R, 2014A&A...564A.122J, 2014ApJ...789L..41J, 2017ApJ...835...50J}, black points and iii) 
  \citet{ 2008ApJ...689.1020Y, 2013ApJ...763...22D, 2013ApJ...765...51L, 2019ApJ...876...43G}, red points.

The lone thin green curve shows only the contribution from AGB stars.
As one can see by comparing this curve with the dashed curves, 
the $\nu$-process only dominates at [Fe/H] $\lesssim -2.2$ for the low yields,
and $\lesssim -1.5$ for the high yields. 
The thicker solid curves show the evolution including AGB and the 3 cases shown by the dashed curves.
As one can see, when the AGB contribution is included, the evolutionary curves fit the data quite
well at all metallicities up to and including solar values. Because of the dispersion at solar metallicity,
it is not possible to distinguish the $\nu$-process contributions at [Fe/H] = 0. 

\begin{figure}
\begin{center}
\epsfig{file=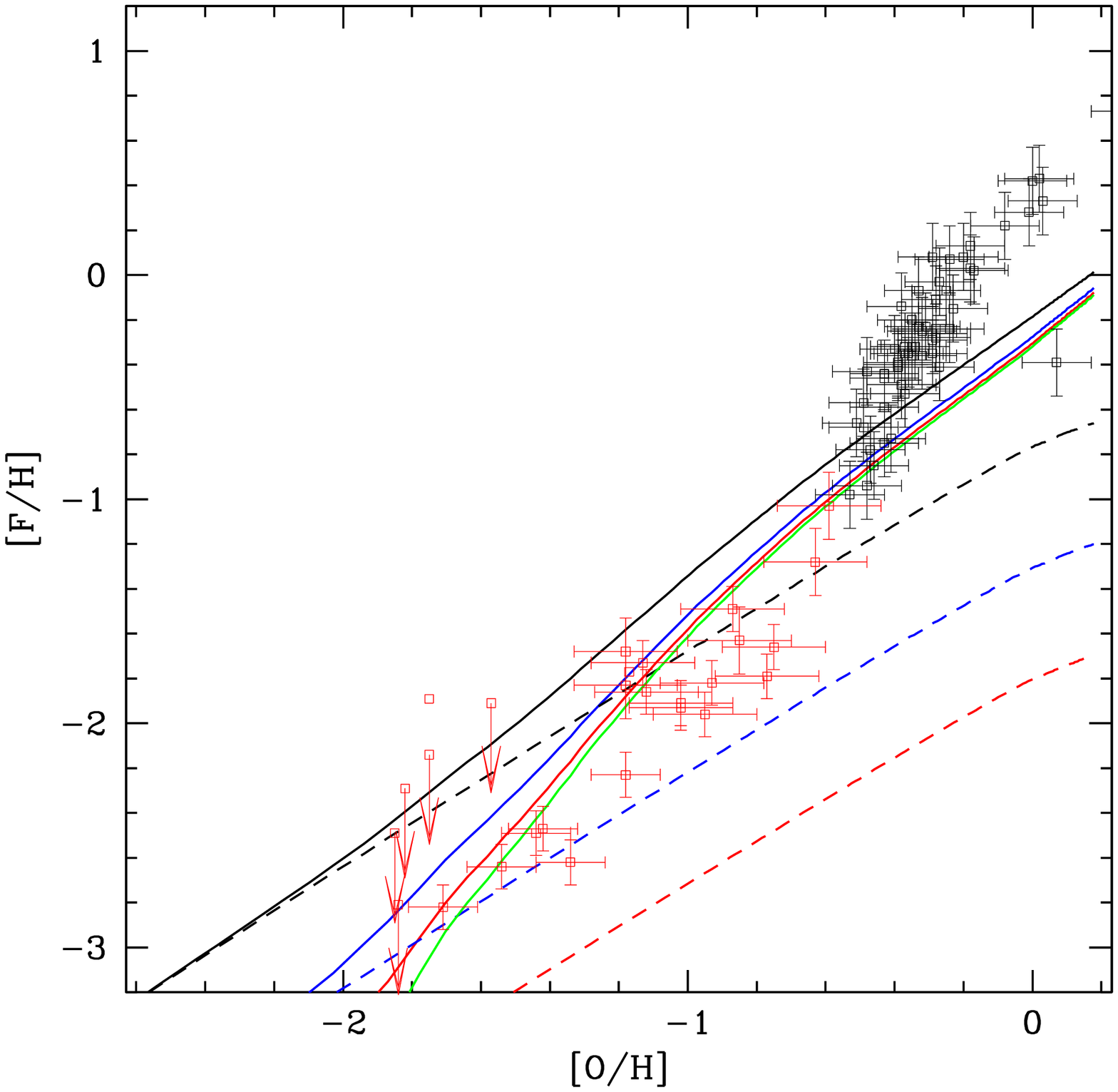, height=3in} \\
\epsfig{file=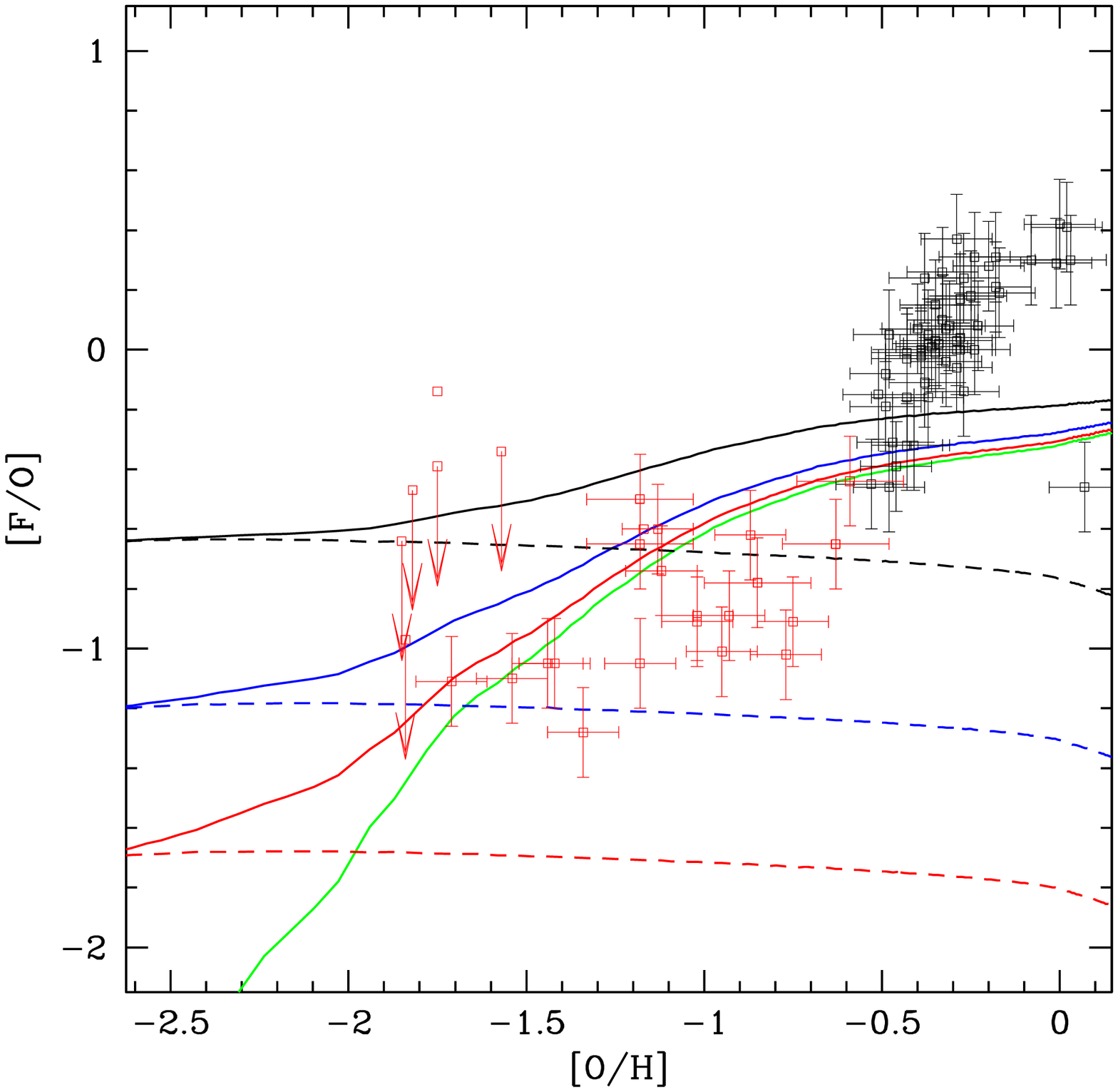, height=3in}
\end{center}
\caption{As in Fig. \ref{fig:FFea}, showing the evolution of [F/H] as a function of [O/H] (upper panel)
and the evolution of [F/O] vs [O/H] (lower panel). }
\label{fig:FOa}
\end{figure}

In  Fig. \ref{fig:FOa}, we show the evolution of fluorine with respect to [O/H]. 
In the upper panel, we plot [F/H[ vs. [O/H] and in the lower panel, 
we plot [F/O] vs. [O/H]. In both panels, we see that the upper limits at low
metallicity are respected, and the curves pass through much of the data at [O/H] $\simeq -1$,
though there is considerable scatter in the data. 
At near solar metallicity, it would appear the models fall well short of the observed data. 
However if we recall the position of this data in the [O/H] vs [Fe/H] plot shown in Fig. \ref{fig:OFe}, we see that these 
points are deficient in oxygen relative to their iron abundance. For this reason, the evolution with respect to 
[Fe/H] fits the data, and the data show low oxygen with respect to the model
prediction for a given fluorine abundance. The data in Fig. \ref{fig:FOa} are taken from i) \citet{2012A&A...538A.117R, 2014A&A...564A.122J, 2014ApJ...789L..41J, 2017ApJ...835...50J} black points and ii) \citet{2008ApJ...689.1020Y, 2013ApJ...763...22D, 2013ApJ...765...51L, 2019ApJ...876...43G} red points.

As in Fig. \ref{fig:FFea}, the dashed show the evolution neglecting the AGB contribution (which alone
is shown by the thin green line). 
We see that the $\nu$-process dominates only for [O/H] ] $\lesssim -1.7$ for the low yields,
and $\lesssim -1.1$ for the high yields. 
The solid curves in  Fig. \ref{fig:FFea} show the fluorine evolution with AGB and $\nu$-process contributions included. 
Once again, at low metallicity, the fluorine abundance is dominated by supernovae (and the $\nu$-process.). 
Our conclusion is that the present fluorine abundance is heavily dominated by AGB production
which minimizes the difference between the high and low energy $\nu$-process yields.

\begin{figure}
\begin{center}
\epsfig{file=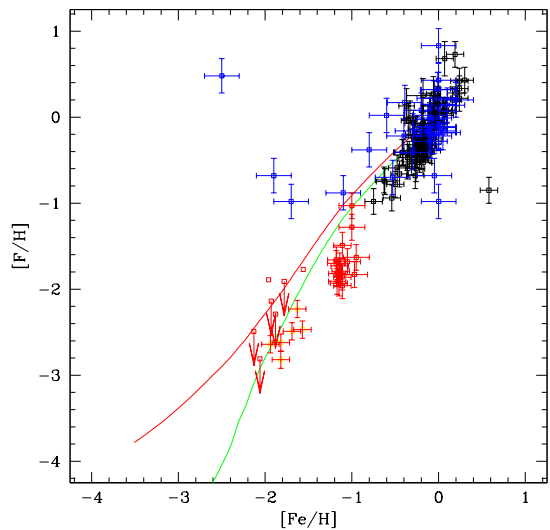, height=3in}
\end{center}
\caption{As in Fig. \ref{fig:BeBFeb} for the evolution of [F/H] as a function of [Fe/H]. All contributions are included
and the $\nu$-process contribution has been scaled to produce the correct \b11/\b10 ratio. }
\label{fig:FFeb}
\end{figure}

In Fig. \ref{fig:FFeb}, we show our final result for the evolution of F.
The neutrino process yield is normalized with $\alpha  = 0.9$ to fix
the \b11/\b10 ratio, and hence there are no additional normalization
adjustments possible for the evolution of F. 
With the exception of some upward scatter at low metallicity, the evolution
which includes both AGB and $\nu$-process contributions fits the data well.

\section{Conclusion}
\label{sec:conclusion}

The relatively low abundances of fluorine make it difficult to observe, and at the same time,
make it difficult to firmly establish the source of fluorine production. 
We have considered three sources for fluorine. The production of F in intermediate mass
stars in the AGB phase, the production in massive stars, and additional 
production in massive stars due to the neutrino process during supernovae. 
The $\nu$-process  contribution always dominates the massive star yield of F
assuming that the low energy yields of \citet{2018ApJ...865..143S} are minimal.

We can resolve the current uncertainty in the $\nu$-process yields by
recognizing the important contribution of this process to the production of \b11 \citep{1994ApJ...424..666O,1996ApJ...468..199V}. 
To this end, we have incorporated the $\nu$ process in a chemical evolution 
model which follows GCRN. The latter explains well the abundance and evolution of Be.
However, the B/Be ratio as well as the isotopic \b11/\b10 ratio are both predicted to be smaller
than observed values. Remarkably, the $\nu$-process production of \b11 provides a substantial
increase in both ratios. Playing on the uncertainty of the $\nu$-process yields,
we are able to fit the \b11/\b10 ratio, with yields slightly above the low energy yields derived in 
\citet{2018ApJ...865..143S}. Once fixed, the B/Be is fully predicted as is the the abundance
and evolution of F. 

While the $\nu$-process is the dominant production mechanism at low metallicity,
the present-day abundance of F, it almost entirely a result of AGB production. 
Even using the higher energy yield neutrino yields would make a marginal increase in 
total present day F abundance. 

Our understanding of the evolution of F or any of the element abundances clearly relies heavily on observation.
Our results show it is clear that the late abundances of F are dominated by AGB production. Thus the
theoretical yields in intermediate mass stars are of prime importance. The potential importance of 
the $\nu$-process in the production of F can only be ascertained with future abundance measurements of F,
particularly at low metallicity.

\section*{Acknowledgements}
This work is made in the ILP LABEX (under reference ANR-10-LABX-63) supported by French state funds managed by the ANR within the Investissements d'Avenir programme under reference ANR-11-IDEX-0004-02. 
The work of  K.A.O. was supported in part by DOE grant DE--SC0011842 at the University of Minnesota. 

\bibliographystyle{mnras}
\bibliography{GCRF}
\label{lastpage}
\end{document}